\documentstyle[12pt]{article}
\begin{document}
\title{LOGICO-ALGEBRAIC APPROACH TO SPACETIME QUANTIZATION}

\author{R.R. Zapatrin,\\
Friedmann Laboratory for Theoretical Physics,\\
SPb UEF, Griboyedova 30/32,\\
191023, St-Petersburg, Russia}

\maketitle

\abstract{ In  the  last  decades  the  logico-algebraic  approach   to
quantum  mechanics  turned  to  be  a  successful tool to render the
quantum   mechanical   formalism   on   a   steady  operationalistic
background [7]. The  algebraic  approach  to  general  relativity  first
proposed by Geroch [1] is used to build a smearing procedure for events
in the quantized theory of spacetime. It  stems
from  the  notion  of  basic   algebra  which  possesses  both   the
differential  structure  (as  functional  algebras  in  [1])   and
non-commutativity (as algebras of observables in quantum  theories).
The main essence  of this formalism  is that it  deprives the notion
of spacetime as  event support of  its fundamental status  making it
relative to  the measurement  performed. Before  the observation  is
executed,  it  is  pointless  to  speak  of  events  as  points   of
spacetime:  there   is  no   underlying  manifold.   Only  when   an
eigen-subalgebra of the basic  algebra is outlined as  eigenstate of
the observable, it  gives rise to  its functional representation  on
the appropriate set thought of as spacetime.
}

\medskip
\medskip
\section{INTRODUCTION}
\medskip
\medskip

In general relativity an event in spacetime is idealized  to  a
point of a four-dimensional manifold. Such idealization is  adequate
within  classical  physics,   but   is   unsatisfactory   from   the
operationalistic point of view. In quantum theory the influence of a
measuring  apparatus  on  the  object  being  observed  can  not  in
principle be removed. We could expect  the  metric  of  a  quantized
theory to be subject to fluctuations, whereas the  primary  tool  to
separate individual events is  just  the  metric [1].  Thus  a  sort  of
{\it smearing procedure}  for events is to be imposed into the theory.

\medskip
An essential step in this direction was the idea to  build  the
differential geometry in terms of abstract algebras.  Geroch  [1]
proposed to generalize the notion of algebra of smooth functions  on
a manifold to that of {\it Einstein algebra} whose elements  are  not  yet
functions. This  generalization  was  successful  since  the  entire
content of general relativity can be reformulated in such a way that
the underlying spacetime manifold is used only once: to  define  the
collection of smooth functions.

Consider the ideas of Geroch in more detail. The basis of the differential
geometry is the notion of {\it vector
field}. It is known any vector field $v$ can  be  associated  with  the
differential operator in the algebra ${\cal A}$ of smooth  functions  on  the
manifold acting as the derivation  along  this  vector  field.  This
operator $v$ is linear, and its main feature is the {\it Leibniz rule}:

$$
v(ab)=v(a)b+av(b)
\eqno{(1)}
$$

\noindent It is  known  that  linear  operators  in ${\cal A}$
satisfying  (1)  are exhausted by that induced by actions of
vector fields. That  is  why the difference is  not  drawn  between
such  operators  and  vector fields: this is  the  essence  of  the
coordinate-free  account  of differential geometry. As a matter of
fact, coordinates appear  only once: to specify the algebra ${\cal A}$
of smooth functions, since the notion of smoothness is referred
to local  maps.  The  forthcoming  notions such as connection,
torsion, curvature  and  others  need  no  local coordinates in
their definition.

\medskip
I emphasize  that  at  the  mere  level  of  definitions  the
principle notions of differential geometry require  no  coordinates,
nor even {\it points}: the fact that ${\cal A}$ is the algebra of  functions
on  a
set is never used. Thus the global geometry {\it per se} does not
confine us by set-theoretical concept of space.

\medskip
\medskip
Although, since the {\it commutative} case is considered, the absence
of points is, roughly speaking, an illusion. As a matter of fact,  a
{\it commutative} algebra can always be represented  by  functions  on  an
underlying space. Such representation is, for instance, the Gel'fand
construction (for normed algebra)  which  is  the  special  case  of
representation of commutative algebra on its spectrum.  So,  in  the
case of the commutative algebra points implicitly exist.

\medskip
The standpoint of the suggested approach is to essentially remove points from
the theory. This  happens  automatically  when  we  pass  to
non-commutative  Einstein  algebras.  Whereas  the  reproduction  of
geometrical constructions causes a  number  of  purely  mathematical
obstacles [2,3].

\medskip
\section{NONCOMMUTATIVITY AND THE SPATIALIZATION PROCEDURE}

\medskip
\medskip
In this section the obstacles arising  in  the
non-commutative generalization  of  the  algebraic  construction  of
differential geometry are overviewed giving rise to some new concepts such
as the {\it spatialization procedure}.

\medskip
\noindent {\bf Basic algebra.} The first question is  why
non-commutativity is to be fetched to geometry? The rough answer is  that  we
have to follow
the tradition of quantization. An amount of non-commutativity in the
geometry  itself  is  needed  to  quantize  it.  This  produces  the
following problem: the lack  of  points  in  this  quantum  geometry
requires a "spatialization" procedure to be imposed into the general
scheme to describe the {\it observable} entities.

\medskip
So, by {\bf basic algebra} of the model I shall mean an  associative  and,
in generally, non-commutative algebra ${\cal A}$ over real (or complex)
numbers which will play the role analogous to that of the algebra of smooth
functions.

\medskip
\noindent {\bf Spatialization procedure}. Let us try to extract
the  geometry  from the basic algebra ${\cal A}$ on its coarsest
level, that is,  set-theoretical one. As it is usually done, we
must consider the elements  of ${\cal A}$  as functions defined on
a certain set $M$, and perhaps, taking the values in a
non-commutative domain {\it R}. That is, the representation of
${\cal A}$  by means of homomorphism $\hat{ }$ is introduced:

$$
a\mapsto\hat{a}
$$

\noindent where $\hat{a}$ is a function $M\rightarrow R$. Thus each
point $m\in M$  is  associated  with the two-sided ideal
$I(m)\subseteq {\cal A}$:

$$
I(m) = \{a\in {\cal A}\mid\hat{a}(m)=0\}
$$

Now we see that the resources of spatialization
are bounded  by the number  of  two-sided  ideals  in ${\cal A}$.
Whereas,  if ${\cal A}$  contains two-sided ideals, it can be, as a
rule,  decomposed  into  mutually commuting components. So, each
point can be associated with at least a simple component of the
decomposition of ${\cal A}$. The conclusion is that {\it
spatialization   and   non-commutativity   are   in    some
sense complementary}: commutation relations can not be described
in  terms of points.

\medskip
When the basic algebra ${\cal A}$  is  commutative  and  satisfies
some additional requirements   (is   Banach   algebra),  the
proposed construction is just the Gel'fand representation endowing
the set $M$ by a natural topology. So, the commutative case makes
it possible to store the {\it topological} space $M$ so that ${\cal
A}$ is represented by continuous functions on {\it M}. However, the
Gel'fand construction  does  not  yield the differential structure
for {\it M}.

Although, the lack of points is  not
an  obstacle  to introduce differential structure with all its
attributes. As in  the commutative case, it is introduced in terms
of the  collection  ${\rm Der}{\cal A}$ of derivations of the basic
algebra ${\cal A}$ \footnote{recall that a {\it derivation} of ${\cal A}$ is
the linear mapping $v:{\cal A}\rightarrow {\cal A}$ enjoying the
Leibniz rule  (1)}.

On the coarse-grained level spacetime is replaced by its finitary
substitute such as {\it Regge space} [4] to restore the metric or {\it
pattern space} [5] to restore only the topology. In the latter case
the points of the space can be restored by a purely algebraic
construction [6].

\medskip
\noindent {\bf Scalars.} In the commutative case we can multiply
a  vector  by  any element of the basic  algebra ${\cal A}$.  In
general,  an  element $v\in {\rm Der}{\cal A}$ multiplied by an element
$a\in {\cal A}$  does  not  enjoy  the  Leibniz  rule.  However, to
define such object as, say,  connection,  multiplicators are
necessary: they play the role of {\it scalars}. So, in the case of
non-commutative basic algebra the notion of scalar is to be carefully
redefined [3,7].

\medskip
\noindent {\bf Einstein equation.} Now consider the point-free  counterpart
of the Einstein equation . In conventional theory it postulates the
equality between the Einstein tensor depending on geometry only  and
the momentum-energy tensor.

\medskip
To form the left side, the analog of the scalar
curvature $R$ is introduced. In classical geometry  $R$   is   the
contraction of
the contravariant metric tensor with the Ricci tensor.  In
differential algebras  we  have  neither  contraction nor tensors,
but   only operators. Although we have the trace of operators in
our  disposal.  In conventional relativity
the operator  form of the Einstein equation is:

$$
R^{i}_{k}- {1\over 2}R\delta ^{i}_{k} = \kappa T^{i}_{k}
$$

In differential algebras the Einstein equation takes the operator form which
requires the introduction of the momentum  operator ${\cal
T}$,  which acts as follows. Recall that $V$ is interpreted as the
set of  virtual shifts, So, if $v\in V$ is associated with a
shift of the  observer, ${\cal T}v$ yields the energy flow he
observes.

\medskip

\noindent {\bf CONCLUDING REMARKS}

\medskip
\medskip
The models of point-free differential geometries  are  proposed
as pairs $({\cal A},V)$, called {\it differential
algebras}  which  are the non-commutative generalization  of
Einstein  algebras  [1].  The  substantial  feature  of
non-commutativity   is   the discrepancy between the elements of
the basic algebra (analog of the smooth functions) and scalars.

\medskip
Luckily,  the  geometry  of  affine  connection   survives   in
non-commutative differential  algebras,  including  the  notions  of
torsion and curvature (as it was shown in [2]). The  scalar  curvature
can  also  be  defined  under   certain
circumstances. If  it  becomes  possible,  the  operator  analog  of
Einstein equation is introduced.

\medskip
The idea to consider vectors as differential operators  applied
to functional algebras, but defined on a  broader  class  of
spaces than manifolds, called {\it differential spaces}, were used
to  implement the spaces with singularities to general relativity.
Heller  {\it et  al.} [8] have shown that the  reasonable
definition  of  differential structure can be  formulated  in
terms  of  certain  algebra ${\cal A}$  of functions so that even
the analog of Lorentz structure can be introduced. In particular,
when ${\cal A}$ is an algebra
of smooth functions on a manifold, the standard differential
geometry is restored.

\medskip
The approach we suggest can be considered as a  reasonable  way to
quantize the gravity. The main problem arising here  is  to  find
the appropriate representations of the basic  algebras.  What
about the source of basic algebras, this is the  Wheeler's
suggestion  to consider  logic  as  pregeometry  which  could  work
here.  The first  step along these lines was made by  Isham [9]:
the  lattice  of  all topologies over a set was
considered, and the analog of creation and annihilation operators
was suggested. The appropriate algebra  could be taken as basic
one. Moreover, starting from an arbitrary property lattice as
background object, one can  always  build  the  semigroup (called
{\it generating} [10])  whose  annihilator lattice
restore this property lattice. Then  the algebra spanned on this
semigroup could play the role of  the  basic algebra ${\cal A}$.

\medskip
The next step is the  {\it spatialization}  procedure. When a
differential structure $V$ and a metric on it
are  set up, the problem arises to extract usual (i.e. point)
geometry  from the triple $({\cal A},V,g)$.  To  return  to
points,  we  must  consider  a subalgebra ${\cal C}\subseteq {\cal
A}$  such  that ${\cal C}$  would  be  commutative  (to   enable
functional representation) and in some sense concerted with $V$ and
$g$.  We did not yet tackle this problem of {\it eigen-subalgebras}
in  detail, whereas it looks as direct way to reveal events within
our  scheme.  It is noteworthy that whenever the triple $({\cal
A},V,g)$ is set  up,  there still    may exist     several
functionally     representable eigen-subalgebras   associated
with    possibly    non-isomorphic geometries.  That  means  that
{\it the  observed  geometry  depends  on observation} which is in
complete accordance with quantum  mechanical point of view.

\medskip
\noindent {\bf ACKNOWLEDGMENTS}

\medskip
\medskip
The attention to the work  offered  by  Frank Antonsen, A.A.Grib, C.J
Isham and the participants of the IQSA'94 (International Quantum
Structures Association Meeting) is appreciated as well as the
financial support from  ISF (George  Soros  Emergency  Grant).

\medskip
\noindent {\bf REFERENCES}

\medskip
\medskip
\noindent [1] Geroch, R.
Einstein Algebras,
{\it Communications in Mathematical Physics},
{\bf 26},
271,
1972

\noindent [2] G.N.Parfionov, R.R.Zapatrin,
Pointless Spaces in General Relativity,
{\it International Journal of Theoretical Physics},
to appear in v. 33,
1994

\noindent [3] G.N.Parfionov,  R.R.Zapatrin,
Dual structures  in  non-commutative differential algebras,
{\it Okayama Mathematical Journal},
submitted in 1994

\noindent [4] T. Regge,
General relativity without coordinates,
{\it Nuovo Cimento},
{\bf 19},
568,
1961

\noindent [5] R.R. Zapatrin,
Pre-Regge  Calculus:  Topology  Via  Logic,
{\it International Journal of Theoretical Physics},
{\bf 32},
779,
1993

\noindent [6] A.A. Grib, R.R. Zapatrin,
Topologymeter: an example of a physical system which is neither
classical nor quantum,
In: IQSA'94 (International Quantum Structures Association) Meeting,
Prague, August 15-21, 1994

\noindent [7] Hooker C.A., ed.,
{\it Logico-algebraic approach to quantum
mechanics: contemporary consolidation},
Reidel, Dordrecht
(1979)

\noindent [8] Heller M., Multarzinski P., and Sasin W.,
{\it The  algebraic approach to space-time geometry},
Acta Cosmologica (Krakow),
{\bf 16},
53
(1989)

\noindent [9] Isham C.J.,
Topology Lattice And The Quantization On The Lattice Of Topologies,
{\it Classical and Quantum Gravity},
{\bf 6},
1509,
(1989)

\noindent [10] Zapatrin R.R.
Quantum Logic Without  Negation,
{\it Helvetica  Physica  Acta},
{\bf 67},
177,
(1994)

\end{document}